\documentclass[journal,twocolumn]{IEEEtran}

\ifCLASSINFOpdf
  \usepackage[pdftex]{graphicx}
  \graphicspath{{fig/}}
  \DeclareGraphicsExtensions{.pdf,.jpeg,.png}
\else
  \usepackage[dvips]{graphicx}
  \graphicspath{{fig/}}
\fi
\ifCLASSOPTIONcompsoc
  \usepackage[caption=false,font=normalsize,labelfont=sf,textfont=sf]{subfig}
\else
  \usepackage[caption=false,font=footnotesize]{subfig}
\fi

\usepackage[cmex10]{amsmath}
\usepackage{amsfonts}
\usepackage{amssymb}
\usepackage{algorithm}
\usepackage{algpseudocode}
\usepackage{subfig}
\usepackage{color,soul}
\usepackage{url}
\usepackage{epstopdf}
\usepackage{float}
\usepackage{amssymb}
\usepackage{enumitem}
\usepackage[compact]{titlesec}
\titlespacing{\section}{0pt}{1ex}{1ex}
\titlespacing{\subsection}{0pt}{1ex}{0ex}
\titlespacing{\subsubsection}{0pt}{1ex}{0ex}
\usepackage{gensymb}
\usepackage{mwe}
\usepackage[export]{adjustbox}
\usepackage{booktabs}
\usepackage{xcolor,colortbl}
\usepackage{bm}
\usepackage{todonotes}
\usepackage{comment} 
\usepackage{multirow}

\begin{document}

\title{A Deep Learning-based Integrated Framework for Quality-aware Undersampled Cine Cardiac MRI Reconstruction and Analysis}

\IEEEaftertitletext{\vspace{-2\baselineskip}}

\author{In\^{e}s P. Machado$^*$, Esther Puyol-Ant\'on, Kerstin Hammernik, Gast\~{a}o Cruz, Devran Ugurlu,  Ihsane Olakorede, Ilkay Oksuz, Bram Ruijsink, Miguel Castelo-Branco, Alistair A. Young, Claudia Prieto, Julia A. Schnabel and Andrew P. King  

\thanks{This work was funded by the Engineering and Physical Sciences Research Council (EPSRC) programme grant ‘SmartHeart’ (EP/P001009/1) and supported by the Wellcome/EPSRC Centre for Medical Engineering [WT 203148/Z/16/Z]. The research was supported by the National Institute for Health Research (NIHR) Biomedical Research Centre and Cardiovascular MedTech Co-operative based at Guy's and St Thomas' NHS Foundation Trust and King's College London. This work was also supported by Health Data Research UK, an initiative funded by UK Research and Innovation, Department of Health and Social Care (England) and the devolved administrations, and leading medical research charities. I. Oksuz has been supported by 2232 International Fellowship for Outstanding Researchers Program of TUBITAK (Project No: 118C353). This research has been conducted using the UK Biobank resource under the application number 17806.} \thanks{I. P. Machado, E. Puyol-Ant\'on, G. Cruz, D. Ugurlu, I. Olakorede, B. Ruijsink, A. A. Young and A. P. King are with School of Biomedical Engineering and Imaging Sciences, King's College London, London, UK. (E-mail:ines.machado@kcl.ac.uk; esther.puyol\_anton@kcl.ac.uk; gastao.cruz@kcl.ac.uk; devran.ugurlu@kcl.ac.uk; ihsane.olakorede@kcl.ac.uk; jacobus.ruijsink@kcl.ac.uk; alistair.young@kcl.ac.uk and andrew.king@kcl.ac.uk)}

\thanks{K. Hammernik is with Technical University of Munich, Munich, Germany and the Biomedical Image Analysis Group, Imperial College London, London, UK. (E-mail:k.hammernik@tum.de)}

\thanks{I. Oksuz is with Istanbul Technical University, Istanbul, Turkey. (E-mail:oksuzilkay@itu.edu.tr)}

\thanks{B. Ruijsink is with Department of Adult and Paediatric Cardiology, Guy’s and St Thomas’ NHS Foundation Trust, London, UK. (E-mail:jacobus.ruijsink@kcl.ac.uk)}

\thanks{M. Castelo-Branco is with Coimbra Institute for Biomedical Imaging and Translational Research (CIBIT/ICNAS), University of Coimbra, Coimbra, Portugal. (E-mail:mcbranco@fmed.uc.pt)}

\thanks{C. Prieto is with School of Biomedical Engineering and Imaging Sciences, King's College London, London, UK and the School of Engineering, Pontificia Universidad Católica de Chile, Santiago, Chile. (E-mail:claudia.prieto@kcl.ac.uk)}

\thanks{J. A. Schnabel is with School of Biomedical Engineering and Imaging Sciences, King's College London, London, UK; Technical University of Munich, Munich, Germany and Helmholtz Center Munich, Munich, Germany. (E-mail:julia.schnabel@kcl.ac.uk)}
}
\maketitle


\begin{abstract}

\boldmath Cine cardiac magnetic resonance (CMR) imaging is considered the gold standard for cardiac function evaluation. However, cine CMR acquisition is inherently slow and in recent decades considerable effort has been put into accelerating scan times without compromising image quality or the accuracy of derived results. In this paper, we present a fully-automated, quality-controlled integrated framework for reconstruction, segmentation and downstream analysis of undersampled cine CMR data.
The framework enables active acquisition of radial k-space data, in which acquisition can be stopped as soon as acquired data are sufficient to produce high quality reconstructions and segmentations. This results in reduced scan times and automated analysis, enabling robust and accurate estimation of functional biomarkers. To demonstrate the feasibility of the proposed approach, we perform realistic simulations of radial k-space acquisitions on a dataset of subjects from the UK Biobank and present results on in-vivo cine CMR k-space data collected from healthy subjects. The results demonstrate that our method can produce quality-controlled images in a mean scan time reduced from 12 to 4 seconds per slice, and that image quality is sufficient to allow clinically relevant parameters to be automatically estimated to within 5\% mean absolute difference. 
\end{abstract}

\ifCLASSOPTIONpeerreview
\else
\fi
\IEEEpeerreviewmaketitle

\section{Introduction}
\label{sec:introduction}

\IEEEPARstart{C}{ardiac} magnetic resonance (CMR) imaging is a common imaging modality for assessing cardiovascular diseases, which are the leading cause of death globally. A complete assessment of cardiac function requires images acquired over the full cardiac cycle. Although cine CMR sequences can provide these images, the acquisition process is slow. To address this, there has been significant research into accelerating acquisitions without compromising the high resolution and image quality offered by cine CMR. One promising approach is to reduce the amount of k-space data acquired. However, cine CMR reconstruction from undersampled k-space data is challenging and state-of-the-art approaches for undersampled reconstruction, such as Parallel Imaging (PI) and Compressed Sensing (CS), are computationally demanding, precluding real-time implementation. Recently, deep learning (DL) based reconstruction approaches have been proposed that involve the acquired k-space data and the forward model directly in the reconstruction network, achieving fast and efficient reconstruction \cite{schlemper2017deep,hammernik2018learning}. However, methods typically do not consider the effect on derived results such as segmentation quality and quantification accuracy.

DL has also significantly impacted downstream processing of cine CMR data. Previous studies have demonstrated automated models for multi-structure segmentation, with performances matching those of human annotators \cite{bai2018automated}. Others have combined DL-based segmentation with automated analysis of volume curves to estimate a range of functional biomarkers \cite{ruijsink2020fully}. However, the majority of these models rely on fully-sampled Cartesian data and thus require lengthy data acquisition. An alternative approach is segmentation estimation directly from undersampled k-space data \cite{schlemper2018cardiac} but bypassing the reconstructed images in this way raises questions of interpretability and clinician trust, and presupposes that the images themselves are not needed for clinical purposes.

Traditionally, cine CMR acquisition, reconstruction and analysis have been considered as independent steps, despite the obvious inter-dependence between high-quality image reconstruction and high accuracy in downstream tasks, such as segmentation and quantification \cite{Oksuz2019b}.
Some preliminary work has combined the reconstruction and segmentation processes in a joint DL-based framework \cite{oksuz2020deep}. However, this was aimed at detecting and correcting for imaging artefacts, not at speeding up the acquisition. Furthermore, quality control (QC) of downstream analysis is an essential component of a clinically-applicable CMR pipeline. Ideally, this QC should be performed whilst the patient is still in the scanner, so that a new scan could be acquired if the original scan was not of sufficient quality. Existing work has not addressed the incorporation of QC into an integrated reconstruction, segmentation and analysis framework.

In this work, we propose the first DL-based framework that integrates cine CMR acquisition, reconstruction and downstream analysis with QC, ensuring that the model outputs are of diagnostic quality. This combination facilitates an active acquisition process, in which only sufficient k-space data are acquired to enable the reconstruction of images that can pass automated QC checks and produce reliable estimates of cardiac functional parameters. The framework aims at fast analysis of undersampled cine CMR data, not only accelerating typically lengthy acquisition times, but also enabling real-time quantification from the reconstructed images.

The remainder of this paper is organised as follows: Section II presents a literature review and our novel contributions in this context. Section III describes the clinical datasets used. Section IV describes our proposed framework for QC-driven reconstruction and analysis of undersampled cine CMR k-space data, including descriptions of each section of the framework. Experiments and Results are presented in Section V and Section VI, respectively, while Section VII discusses the findings of this paper in the context of the literature and suggests potential directions for future work.


\section{Related work}
\label{sec:Related}

 
\subsection{Acquisition and reconstruction} 
\label{sec:reconstruction}
\vspace{0.1cm}
Considerable effort has been devoted to accelerate the reconstruction of cine CMR from undersampled k-space data including PI \cite{uecker2014espirit} and CS \cite{menchon2019reconstruction}. CS approaches work by exploiting redundancy or assumptions about the underlying data to resolve the aliasing caused by sub-Nyquist sampling. CS is computationally demanding, motivating recent research into learning the reconstruction mapping from k-space data to reconstructed images via convolutional neural networks (CNNs) \cite{kustner2020cinenet,qin2018convolutional}. It has already been demonstrated that CNNs outperform sparsity-based methods with respect to both reconstruction quality and speed \cite{shi2016real}, making clinical deployment feasible. DL solutions for CMR reconstruction in general can be classified as those that mimic the optimization process of iterative reconstruction approaches by unrolling the process into several stages, such as \cite{schlemper2017deep} and \cite{hammernik2018learning}, and those that pursue reconstruction as a black-box model, such as \cite{zhu2018image} and \cite{lee2018deep}. For instance, the method proposed in \cite{schlemper2017deep} describes a deep neural network trained to reconstruct cine sequences of CMR images. This results in an iterative procedure consisting of a cascade of two structures, a Deep Network (DN) and a Data Consistency (DC) unit. More recently, DL approaches have been proposed by exploiting spatio-temporal redundancy via recurrent CNNs \cite{qin2018convolutional}. K\"{u}stner et al. proposed the CINENet network for 3D+time cine CMR reconstruction and showed that it outperforms iterative reconstruction in terms of visual image quality and contrast \cite{kustner2020cinenet}. These works clearly show the benefits of the proposed architectures and are now starting to be deployed into MR scanner software \cite{jaubert2021real}. 

\subsection{Segmentation and quantification}
\label{sec:segmentation_quantification}
\vspace{0.1cm}
Image segmentation is an important downstream task for many cardiovascular clinical applications. Segmentation enables the quantification of parameters that describe cardiac morphology, such as left ventricle (LV) and right ventricle (RV) end-diastolic (ED) and end-systolic (ES) volumes, or cardiac function, e.g. myocardial wall thickening and ejection fraction (EF). 
A large body of research has been dedicated to developing automated cine CMR segmentation methods \cite{peng2016review}.
Many such methods are based on the U-Net architecture \cite{ronneberger2015u}. For instance, a basic CNN architecture with 9 convolutional layers and a single up-sampling layer was used to segment short-axis (SAX) cine CMR images \cite{romaguera2018myocardial}. Another example is a fully convolutional approach with a simpler up-sampling path that was successfully applied for pixel-wise segmentation of 4-chamber, 2-chamber and SAX cine CMR images in less than 1 min  \cite{bai2018automated}. More recently, the nnU-Net framework \cite{isensee2021nnu} has shown state-of-the-art performance for automatic segmentation of both ventricles and the myocardium from cine CMR \cite{Mariscal2021}. nnU-net was the top-performing model in the Automated Cardiac Diagnosis Challenge (ACDC) CMR segmentation challenge \cite{bernard2018deep}. Once accurate cine CMR segmentations have been produced, morphological and functional parameters can be calculated from the segmentations. Frameworks have been proposed for estimating LV and RV volumes and EF \cite{attar2019quantitative}, with others going further by estimating parameters from the atria and aorta \cite{bai2020population} as well as a wider range of systolic and diastolic functional parameters \cite{ruijsink2020fully}. Some of these frameworks \cite{ruijsink2020fully, attar2019quantitative} also incorporate QC checks to enable their use in clinical imaging and retrospective population studies.

\subsection{Quality control}
\label{sec:qualitycontrol}
\vspace{0.1cm}
In the medical imaging domain, quality assessment is an important topic of research in the fields of image acquisition, reconstruction and segmentation. In acquisition and reconstruction, motion during the CMR scanning process is a major source of image degradation \cite{mojibian2022cardiac}.
This can lead to artefacts such as blurring, ghosting, and breath-hold slice misalignment. It is important to be able to automatically detect when a reconstruction method fails, so as to avoid inclusion of wrong measurements into subsequent analyses and potentially incorrect conclusions. 

Work on QC in general MR imaging includes K\"{u}stner et al., who proposed to extract a set of features and trained a deep neural network for artefact detection \cite{kustner2018machine}. Previous work on QC in cine CMR imaging includes Zhang et al. \cite{zhang2016automated}, who proposed a method to identify missing apical and basal slices. More recently, Oksuz et al. \cite{oksuz2019automatic} used a curriculum learning strategy exploiting different levels of k-space corruption to detect cardiac motion artefacts. This work was extended in \cite{oksuz2020deep} to both detect and correct for the motion artefacts. 

In clinical applications, using erroneous segmentations of medical images can have dramatic consequences. Recently, methods have been proposed to detect segmentation failures in large-scale CMR imaging studies for removal from subsequent analyses. Using the approach of Reverse Classification Accuracy (RCA), Robinson et al. \cite{robinson2019automated} predicted cine CMR segmentation metrics to identify failed segmentations, achieving good agreement between predicted metrics and visual QC scores. Galati et al. \cite{galati2021efficient} proposed a convolutional autoencoder to quantify segmentation quality at inference time, without a ground truth (GT). More recently, Fournel et al. \cite{fournel2021medical} proposed a new CNN-based segmentation QC approach by training a DL classifier on CMR images and derived segmentations to predict quality.

\subsection{Contributions}
\label{sec:contributions}
\vspace{0.1cm}
\noindent There are five major contributions of this work:
\begin{itemize}[leftmargin=*]
    \item We propose a DL-based and QC-driven integrated framework which can automatically reconstruct and segment undersampled cine SAX CMR images at all time points across the cardiac cycle and, from these, derive functional biomarkers. 
    \item The framework includes robust pre- and post-analysis QC mechanisms to detect high-quality image reconstructions and segmentations. 
    \item Importantly, the framework enables active acquisition of radial k-space data, in which acquisition can be stopped as soon as acquired data are sufficient to produce high quality reconstructions and segmentations. 
    \item We analyse performance on a set of 16 in-vivo acquisitions to illustrate the potential of our technique for cine CMR image reconstruction and analysis.
    \item We show that quality-controlled cine CMR images can be reconstructed in a scan time reduced from 12 to 4 seconds per slice, and that image quality is sufficient to allow for clinically relevant parameters to be automatically estimated to within 5\% mean absolute error. 
\end{itemize} 

This paper builds upon our previous work on cine CMR reconstruction and analysis \cite{machado2021quality}. Here, we extend this preliminary work in two main ways. First, we include a DL-based segmentation QC step that enables our framework to be run in real-time. This allows, for the first time, the possibility of an active acquisition process in which the quality of reconstructed images and derived segmentations can be used to control the k-space profile acquisition process, resulting in faster acquisitions. Second, we evaluate our active acquisition framework on in-vivo cine CMR k-space data in addition to the realistic simulations we previously performed.

\begin{figure*}[t] 
\centering
\centerline{\includegraphics[width=\linewidth]{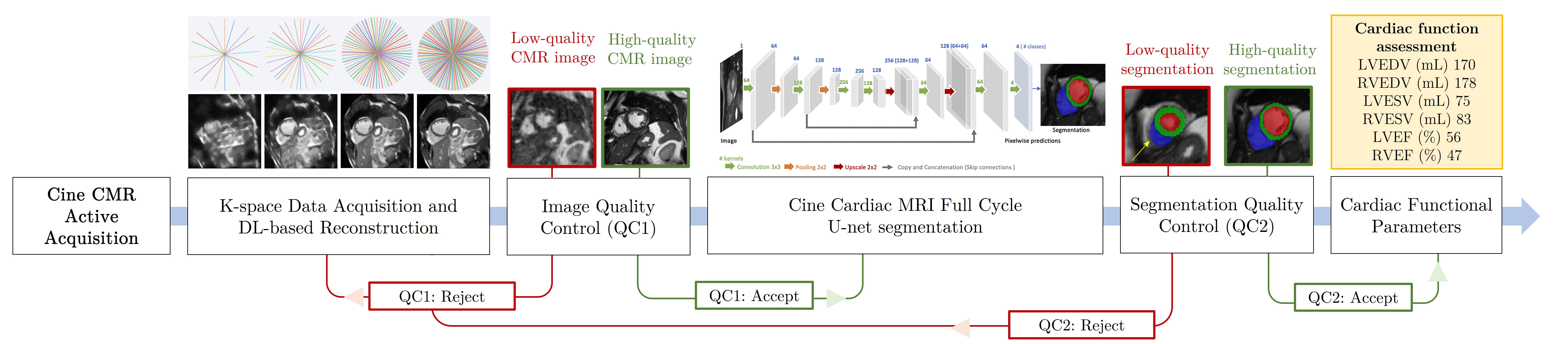}}
\caption{Overview of the image analysis pipeline for fully-automated cine CMR undersampled reconstruction and analysis including comprehensive QC algorithms to detect erroneous output. As k-space profiles are acquired, images are continually reconstructed using a Deep Cascade of Convolutional Neural Networks (DCCNN) and passed through QC checks: QC1 to detect high-quality reconstructions and QC2 to detect high-quality segmentations. The active acquisition terminates when the reconstructed images pass all QC checks.}
\label{fig:pipeline}
\end{figure*}

\vspace{0.1cm}
\section{Materials}
\label{sec:materials}
\vspace{0.1cm}

\subsection{UK Biobank cine CMR imaging data}
\label{sec:ukbbdataset}
\vspace{0.1cm}
The reconstruction and segmentation models were trained using a subset of 4,875 cine SAX CMR scans acquired from healthy subjects from the UK Biobank. The QC models were trained using an additional set of 100 cases from the UK Biobank. The UK Biobank cine CMR scans were all acquired using a 1.5 Tesla MRI scanner (MAGNETOM Aera, Siemens Healthcare, Erlangen, Germany). The SAX image acquisition typically consists of 10 image slices with a field of view of 380 × 252 mm and a slice thickness of 8 mm, covering both ventricles from the base to the apex. The in-plane image resolution is 1.8 × 1.8 mm\textsuperscript{2}, the slice gap is 2 mm, with a repetition time (TR) of 2.6 ms and an echo time (TE) of 1.10 ms. Each cardiac cycle consists of 50 time frames. More details of the image acquisition protocol can be found in \cite{petersen2016uk}. \\
\indent To train the reconstruction and QC models, the UK Biobank data were used to simulate cine CMR images based on undersampled radial k-space trajectories with a golden-angle step (TR = 2.6 ms) containing increasing numbers of profiles corresponding to scan times between 1 to 30 seconds, in steps of 1 second. A more detailed description of the simulated phase and radial acquisition pattern can be found in Sections \ref{sec:simulated_data} and \ref{sec:radial}, respectively. 

To train the segmentation model, pixel-wise segmentations of three structures (LV, RV and myocardium) for both ED and ES frames were manually performed to act as GT segmentations. The segmentations were performed by a group of eight observers and each subject was annotated only once by one observer. Visual QC was performed on a subset of the data to ensure acceptable inter-observer agreement. The segmentation model was evaluated using 600 different subjects (with GT segmentations of the three structures) from the UK Biobank for intra-domain testing and two other datasets for cross-domain testing: the ACDC dataset (100 subjects, 1 site, 2 scanners) and the British Society of Cardiovascular Magnetic Resonance Aortic Stenosis (BSCMR-AS) dataset (599 subjects, 6 sites, 9 scanners). \\
\indent In addition, an extra cohort of healthy (n=200) and cardiomyopathy (n=70) cases from the UK Biobank was used to evaluate the complete framework. These were not used for training/testing/validating any component of the framework.

\subsection{In-vivo cine CMR imaging data}
\label{sec:realspacedata}
\vspace{0.1cm}
To demonstrate the performance of the framework in a more realistic scenario, cine CMR k-space data from 16 healthy subjects was acquired in-vivo using a tiny golden angle radial bSSFP sequence with angular step of $23\degree$ on a 1.5 Tesla MRI scanner (Ingenia, Philips, Best, The Netherlands). Further acquisition parameters are: TR = 2.3 ms, TE = 1.1 ms, in-plane resolution = 2 mm × 2 mm, slice thickness = 8 mm, flip angle = 60° and number of channels = 28. Retrospective ECG-triggering was used to reconstruct 25 cardiac phases in a segmented fashion. The total scan time was 20 heartbeats on average. Ethical approval was obtained and all subjects provided informed consent.

\section{Methods}
\label{sec:methods}

In this section, we describe the integrated framework for active and quality-controlled cine CMR image acquisition, reconstruction and downstream analysis. We provide details of the specific models used for the reconstruction of 2D+time cine SAX CMR from undersampled k-space data (Section \ref{sec:model_reconstruction}), image QC to detect high-quality reconstructions (Section \ref{sec:model_imageQC}), bi-ventricular segmentation (Section \ref{sec:model_segmentation}), a QC step to detect high-quality segmentations (Section \ref{sec:model_segmentationQC}) and automated calculation  of  cardiac functional parameters (Section \ref{sec:model_estimates}). For an illustration of the pipeline see Fig. \ref{fig:pipeline}.

\subsection{Reconstruction}
\label{sec:model_reconstruction}

\begin{figure*}[ht!]
\centering
\centerline{\includegraphics[width=0.75\linewidth]{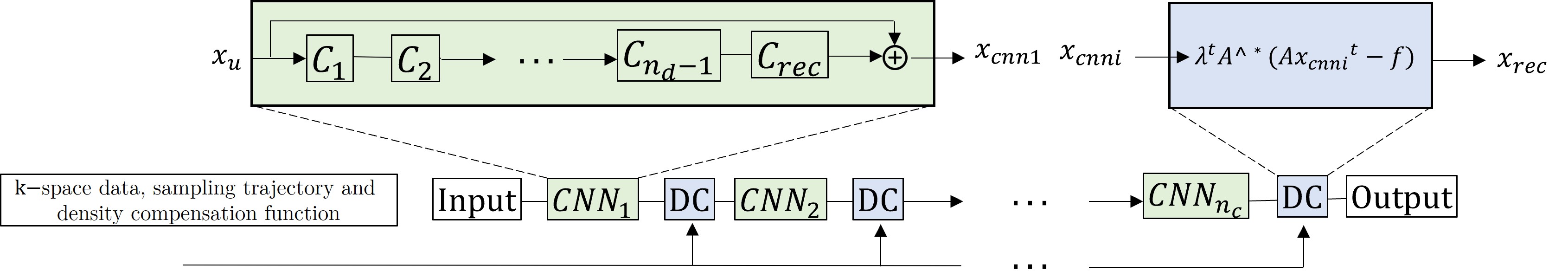}}
\caption{A cascade of CNNs. $x_u$ $\in$ $\mathbb{C}^{N}$ represent a sequence of 2D complex-valued MR images stacked as a column vector, where N = $N_x$ $N_y$ $N_t$. To obtain a reconstruction, we feed the undersampled k-space data, sampling trajectory and density compensation function to the network. CNN and DC denotes the convolutional neural network and the data consistency layer, respectively. The number of convolutional layers $C_i$ within each network and the depth of cascade is denoted by $n_d$ and $n_c$ respectively. The final layer of the CNN module is a convolution layer $C_{rec}$ which projects the extracted representation back to the image domain. The data consistency layers follow a gradient descent scheme, where $A^*$ is the adjoint linear sampling operator, $f$ is the given undersampled k-space data and $t$ is the iteration number, according to \cite{hammernik2018learning}. During the training procedure, the filter kernels, activation functions and data term weights $\lambda^{t}$ are learned.}
\label{fig:dccnn}
\end{figure*}

As k-space profiles are acquired, images are continually reconstructed using the Deep Cascade of Convolutional Neural Networks (DCCNN) method \cite{schlemper2017deep}, a CNN-based framework for reconstructing MR images from undersampled data to accelerate the data acquisition process. DCCNN features alternating data consistency layers and regularisation layers within an unrolled end-to-end framework. Undersampled k-space data, along with the sampling trajectory and density compensation function, are provided as input to this unrolled model for DL reconstruction, and high-quality CMR images are obtained as an output in an end-to-end fashion. The regularisation layers of this network were implemented as a 5-layer CNN according to \cite{schlemper2017deep}, and the data consistency layers follow a gradient descent scheme according to \cite{hammernik2018learning}. The reconstruction model was trained using a subset of 4,875 cine CMR scans acquired from healthy subjects from the UK Biobank (3,975 cases were used for training, 300 for validation and 600 for testing the model). More details of the image reconstruction method can be found in \cite{hammernik2018learning}. The network architecture for the CNN is illustrated in Fig. \ref{fig:dccnn}.

\subsection{Image quality-control network}
\label{sec:model_imageQC}

The first QC step (QC1) was framed as a binary classification problem and addressed using a ResNet classification network \cite{he2016deep}. We chose a residual network because it can make the training process faster and it achieves state-of-the-art performance. For training the ResNet, the UK Biobank data from 100 healthy subjects were used to simulate cine series based on undersampled radial k-space trajectories containing increasing numbers of profiles corresponding to scan times between 1 to 30 seconds, in steps of 1 second. The k-space data were reconstructed using the DCCNN reconstruction model. Slice-level binary image quality labels (analyzable/non-analyzable) were generated by visual inspection from 30,000 2D images (100 subjects * 10 slices * 2 time frames * 15 undersampling factors) at different levels of undersampling. Visual QC was performed by an expert cardiologist on a subset of the data to ensure acceptable inter-observer agreement. Images were considered to be high-quality if they were acquired at the correct slice location, were artefact-free and had good image contrast throughout the cardiac cycle. Due to the imbalance in the labels, we used the precision-recall metric to evaluate classifier output quality. Prior to training, all images were cropped to a standard size of 192 × 192 pixels and 80\% were used for training, 10\% for validation, and 10\% for testing the network. The ResNet was trained for 200 epochs with a binary cross entropy loss function. During training, data augmentation was performed on-the-fly including rotation, shifts and image intensity transformations. The probability of augmentation for each of the parameters was 50\%. The training/validation/testing images for QC1 were randomly selected from the UK Biobank dataset and were not used for training or evaluating the reconstruction/analysis framework.

\subsection{Full cycle image segmentation}
\label{sec:model_segmentation}

\begin{figure*}[t]
\centering
\centerline{\includegraphics[width=0.80\linewidth]{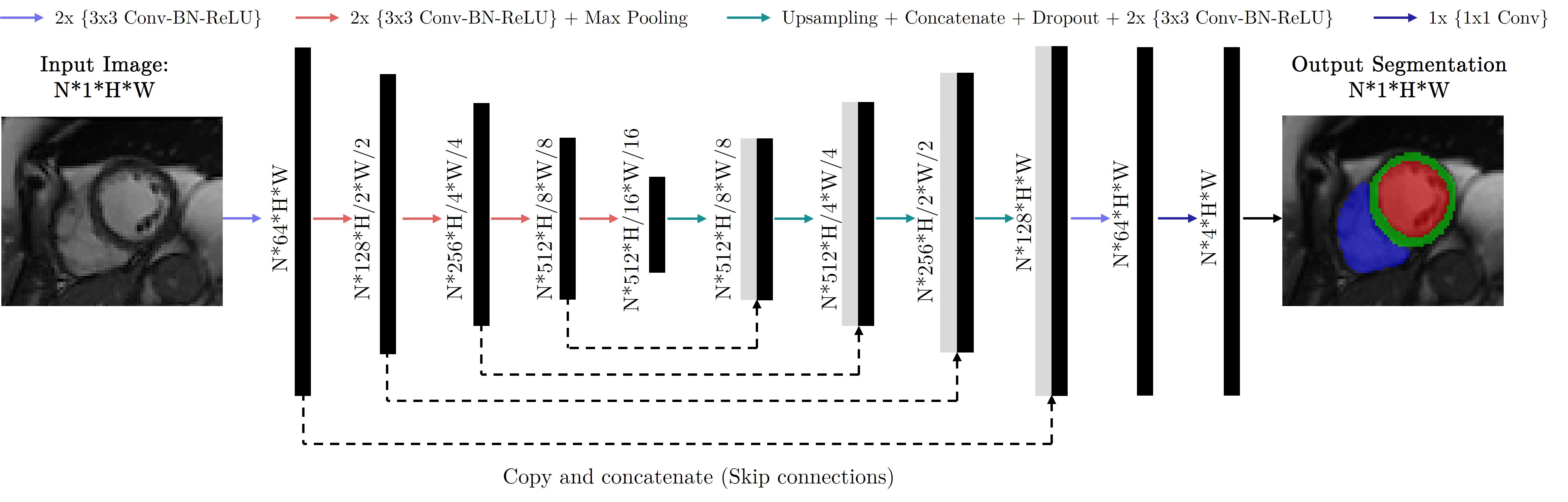}}
\caption{U-net architecture used for segmentation of 2D+time CMR sequences. The U-Net takes a batch size of N 2D CMR images as input at each iteration, learning multi-scale features through a series of convolutional layers and max-pooling operations. These features are then combined through upsampling and convolutional layers from coarse to fine scales, generating pixel-wise predictions for the four classes (background, LV, RV and myocardium) on each slice. Conv: Convolutional layer; BN: Batch normalization; ReLU: Rectified linear unit.}
\label{fig:UnetModel}
\end{figure*}

We used a U-net based architecture for automatic segmentation of the LV blood pool, LV myocardium and RV blood pool from all SAX slices and all frames throughout the cardiac cycle. Fig. \ref{fig:UnetModel} shows the U-net architecture used for segmentation of the cine sequences. The subset of the UK Biobank dataset with GT annotations was split into 3,975, 300 and 600 subjects for training, validation and testing respectively. All images were resampled to 1.25 × 1.25 mm. The training dataset was augmented in order to cover a wide range of geometrical variations in terms of the heart pose and size. During training, the dropout rate for each layer was set to be 0.2. In every iteration, cross entropy loss was calculated to optimize the network parameters through back-propagation. Specifically, the stochastic gradient descent (SGD) method was used during the optimization, with an initial learning rate of 0.001. The learning rate was decreased by a factor of 0.5 every 50 epochs. More details of the image segmentation method can be found in \cite{chen2020improving}.

\subsection{Segmentation quality-control network}
\label{sec:model_segmentationQC}

The second QC step (QC2) was also framed as a binary classification problem and addressed using a ResNet classification network \cite{he2016deep}, which took an image-segmentation pair as input, similar to \cite{fournel2021medical}. To define the binary labels, we first calculate the Dice Similarity Coefficient (DSC) per-class (LV and RV blood pool and LV myocardium) between predicted segmentations and the manually-labelled segmentations. A good quality pair was assumed to have a mean DSC for all classes above 0.7 and the slice-level binary labels were defined accordingly. To train and evaluate QC2, we used a subset of 100 subjects from the UK Biobank. This resulted in a total of 30,000 samples (100 subjects * 10 slices * 2 time frames * 15 undersampling factors). Due to the imbalance in the DSC scores, we chose to sample the DSC distribution using the following bins: [0, 0.2], [0.2, 0.3], [0.3, 0.4], [0.4, 0.5], [0.5, 0.6], and [0.7, 1]. We then took a fixed number of segmentations from each bin, equal to the minimum number of counts-per-bin across the distribution as in \cite{fournel2021medical}. Our final dataset comprised 23,520 samples. We split the data at the subject level 80:10:10 giving 18,816 training samples and 2,352 samples each for validation and testing. The ResNet was trained for 200 epochs with a binary cross entropy loss function and a precision-recall metric to evaluate classifier output quality. During training, data augmentation was performed on-the-fly including rotation, shifts and image intensity transformations. The probability of augmentation for each of the parameters was 50\%. The training/validation/testing images for QC2 were randomly selected from the UK Biobank dataset and were not used for training or evaluating any other parts of the reconstruction/analysis framework.

\subsection{Clinical functional parameters}
\label{sec:model_estimates}

A range of functional biomarkers were derived from the image segmentations. We calculated the left ventricle end-diastolic volume (LVEDV), left ventricle end-systolic volume (LVESV), left ventricle ejection fraction (LVEF), right ventricle end-diastolic volume (RVEDV), right ventricle end-systolic volume (RVESV) and right ventricle ejection fraction (RVEF). The volumes were calculated by multiplying the number of voxels across all slices by the voxel volume for each of the LV/RV classes. The maximum volume over the cardiac cycle was used for (LV/RV)EDV and the minimum for (LV/RV)ESV. EF (for both LV and RV) was calculated as (EDV-ESV)/EDV. 

\subsection{Implementation details}
\label{sec:implementation}
\vspace{0.1cm}
All experiments were performed on a single desktop computer equipped with a quad-core 3.5 GHz CPU, 16 GB RAM and an NVIDIA GTX 1080 Ti GPU. The Pytorch framework was used for implementation.

\section{Experiments}
\label{sec:experiments}

We evaluated our framework using two types of experiments. In Experiment I, we used reconstructed cine CMR images from the UK Biobank in order to generate complex-valued k-space data by simulating a synthetic phase and a golden-angle radial acquisition process, as described in Sections \ref{sec:simulated_data} and \ref{sec:radial}, respectively. In Experiment II, we used in-vivo cine CMR complex images generated as sensitivity-weighted coil-combination and simulated a golden-angle radial acquisition process, as described in Sections \ref{sec:real_data} and \ref{sec:radial}, respectively. Section \ref{sec:recons} describes the reconstruction methods used for comparison. Section \ref{sec:metrics} describes the metrics used to validate image and segmentation quality during active acquisition.

\subsection{Evaluation metrics}
\label{sec:metrics}

Image quality was evaluated with Mean Absolute Error (MAE), Structural Similarity Index (SSIM) and Peak Signal to Noise Ratio (PSNR), defined in Equations \eqref{eq:mae}, \eqref{eq:ssim} and \eqref{eq:psnr}, respectively. 
\begin{equation}
{\text{MAE}}=  \dfrac{1}{N_{p}} \sum_{p=1}^{N_{p}} | (I_{x}(p)-I_{y}(p)) |
\label{eq:mae}
\end{equation}
where $p$ corresponds to each pixel in a total of $N_{p}$ pixels in reference and predicted images $I_{x}$ and $I_{y}$.

\begin{equation}
\text{SSIM}(p)=  \dfrac{(2 \mu_{x}\mu_{y}+c_{1}) (2 \sigma_{xy}+c_{2})}{ (\mu_{x}^{2}+\mu_{y}^{2}+c_{1}) (\sigma_{x}^{2}+\sigma_{y}^{2}+c_{2})}  
\label{eq:ssim}
\end{equation}
where ($\mu_{x}$, $\sigma_{x}$) and ($\mu_{y}$, $\sigma_{y}$) correspond to the average intensities and variance values for regions $x$ and $y$, respectively. $\sigma_{xy}$ is the covariance of regions $x$ and $y$ and $c_{1}$ and $c_{2}$ are constant values for stabilising the denominator. 
\vspace{-0.1cm}
\begin{equation}
{\text{PSNR}}=  20 \log_{10} (\text{max}(I)) - 10 \log_{10}\Big(\dfrac{1}{N_{p}} \sum_{p=1}^{N_{p}}  (I_{x}(p)-I_{y}(p))^2\Big)  
\label{eq:psnr}
\end{equation}
where $ \text{max}(I)$ corresponds to the maximum intensity value in the reference image.

To evaluate the quality of segmentations, the DSC overlap measure was computed. DSC is defined between two regions $A$ and $B$ by Equation \eqref{eq:dsc}.

\begin{equation}
DSC(A,B)=\dfrac{2 \|A \cap B \|}{\|A\| \cup \|B\|}.
\label{eq:dsc}
\end{equation}

DSC is a value between 0 and 1, with 0 denoting no overlap and 1 denoting perfect agreement. The higher the DSC value, the better the agreement.

\subsection{Experiment I: Retrospective undersampling from UK Biobank data with simulated phase}
\label{sec:simulated_data}

Phase information is an important source of data in cine CMR image reconstruction. However, the UK Biobank dataset contains only the reconstructed magnitude images. In this experiment, we use reconstructed cine CMR images from the UK Biobank to produce fully-sampled complex-valued k-space data with synthetic but realistic phase information. More details on the dataset used in this experiment can be found in Section \ref{sec:ukbbdataset}. To generate the synthetic phase, we applied a low pass filter to the k-space data followed by an inverse Fourier transform similar to \cite{Haldar2013}. White Gaussian noise was added to the original images and then a Fourier Transform was applied to generate realistic k-space data. We used these data to simulate an active acquisition process by using a radial golden-angle sampling pattern to produce undersampled k-space with varying undersampling degrees, as described in Section \ref{sec:radial}. The functional metrics estimated from the undersampled reconstructed images that passed the QC checks with the lowest scan time during active acquisition were compared to those derived from the GT segmentations of the fully-sampled images.

\subsection{Experiment II: Retrospective undersampling from in-vivo coil-combined fully-sampled data}
\label{sec:real_data}

In this experiment, we used in-vivo tiny golden-angle radially-acquired cine CMR fully-sampled complex-valued k-space data from 16 healthy subjects. More details on the dataset used in this experiment can be found in Section \ref{sec:realspacedata}. We used these data to simulate an active acquisition process by using a radial golden-angle sampling pattern to produce undersampled k-space with varying undersampling degrees, as described in Section \ref{sec:radial}. The functional metrics estimated from the undersampled reconstructed images that passed the QC checks with the lowest scan time during active acquisition were compared to those derived from the GT segmentations of the fully-sampled images.

\subsection{Simulation of radial acquisition pattern}
\label{sec:radial}

\begin{figure}[t] 
\centering
\centerline{\includegraphics[width=\linewidth]{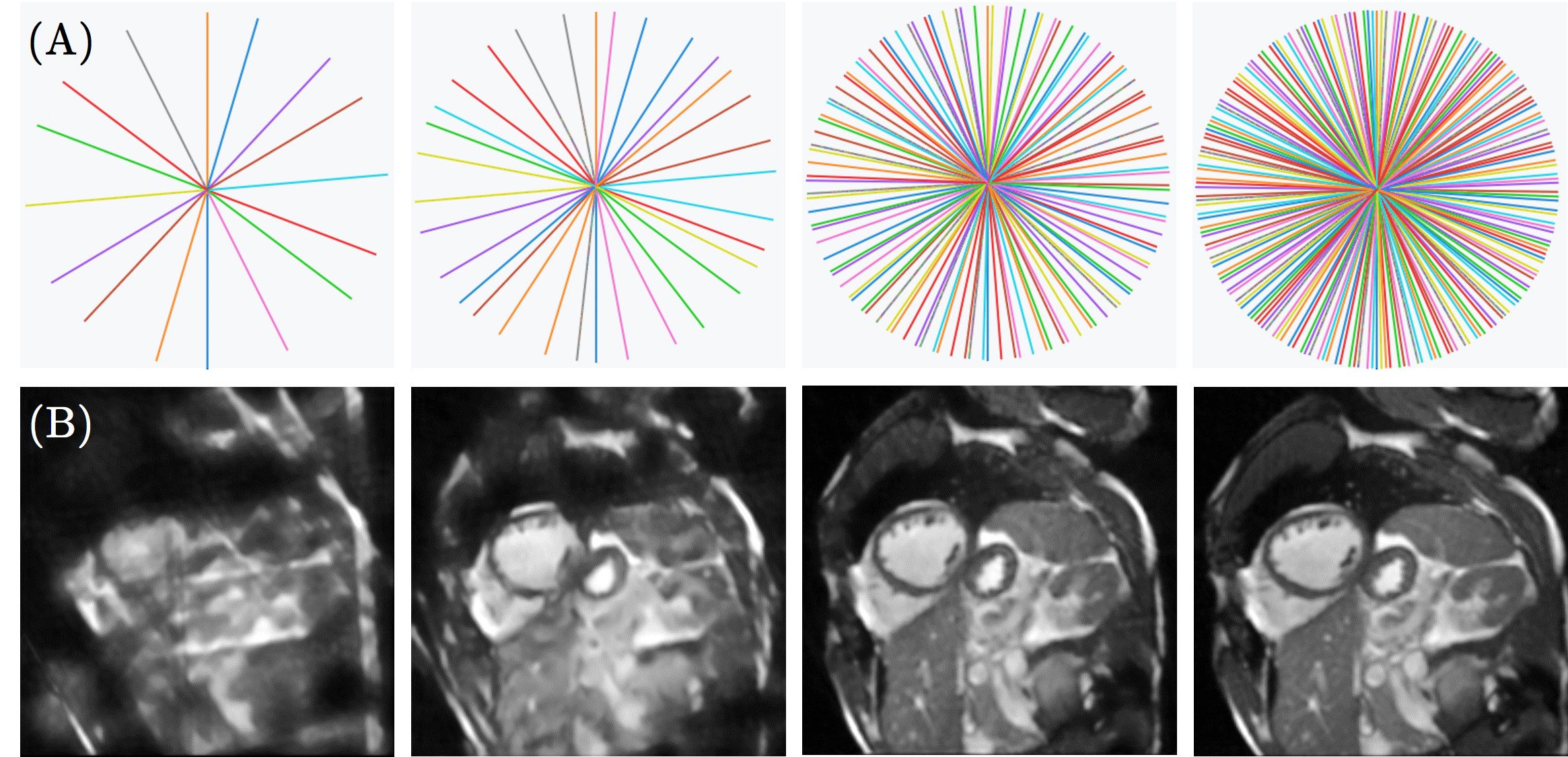}}
\caption{Simulation of radially undersampled data from fully-sampled Cartesian data. Pseudo-radial patterns (A) were used to mask the Cartesian grid with a different number of projections as described in Section \ref{sec:radial}. (B) shows the result of applying the simulated undersampling to the fully-sampled data. After each acquisition of radial profiles, the undersampled k-space data were used as input to our framework.}
\label{fig:undersampling}
\end{figure}

We simulated an active acquisition process by using a radial golden-angle sampling pattern to produce undersampled k-space data from simulated (Experiment I) and in-vivo (Experiment II) fully-sampled reconstructed images. These data contain increasing numbers of profiles corresponding to scan times between 1 to 30 seconds, in steps of 1 second. To simulate the undersampled radial k-space acquisition, the images were organized into 3D matrices. The resulting matrix was Fourier transformed along the spatial domains and each ($k_x$-$k_y$) space was masked by a radial pattern with a TR of 2.6 ms. For each sample, the number of projections per frame (t) was equal to $P \in$ \{7, 15, 23, 30, 38, 46, 53, 61, 69, 77, 84, 92, 100, 107, 115, 123, 130, 138, 146, 153, 161, 169, 176, 184, 192, 200, 207, 215, 223, 230\} corresponding to thirty different sampling rates. Accordingly, the corresponding acceleration factors R with respect to the radial fully-sampled data were \{41.96, 19.58, 12.77, 9.79, 7.73, 6.39, 5.54, 4.82, 4.26, 3.81, 3.5, 3.19, 2.94, 2.74, 2.55, 2.39, 2.26, 2.13, 2.01, 1.92, 1.83, 1.74, 1.67, 1.60, 1.53 ,1.47, 1.42, 1.37, 1.32, 1.28\}. 
The angle increment between projections within one frame was set to the golden angle \cite{winkelmann2006optimal}. The obtained series was transformed back to image space representing reconstructions of undersampled k-spaces, as shown in Fig. \ref{fig:undersampling}.

\subsection{Reconstruction methods}
\label{sec:recons}

As well as our proposed DCCNN reconstruction approach described in Section \ref{sec:model_reconstruction}, we performed a comparative evaluation with reconstruction using a non-uniform Fast Fourier Transform (nuFFT).

\section{Results}
\label{sec:results}

We now present results that illustrate the ability of our integrated framework to produce high-quality image reconstructions and segmentations (Sections \ref{sec:qc1_results} and \ref{sec:qc2_results}, respectively) and accurate estimates of cardiac functional parameters  (Section \ref{sec:functional_results}). Section \ref{sec:scantime} shows differences in terms of scan time and acceleration factors between the fully-sampled acquisition and our integrated framework. A qualitative assessment of the framework was also performed and results can be found in Section \ref{sec:qualitative}. 

\subsection{Image quality analysis}
\label{sec:qc1_results}

\begin{figure*}[t] 
\centering
\centerline{\includegraphics[width=0.70\linewidth]{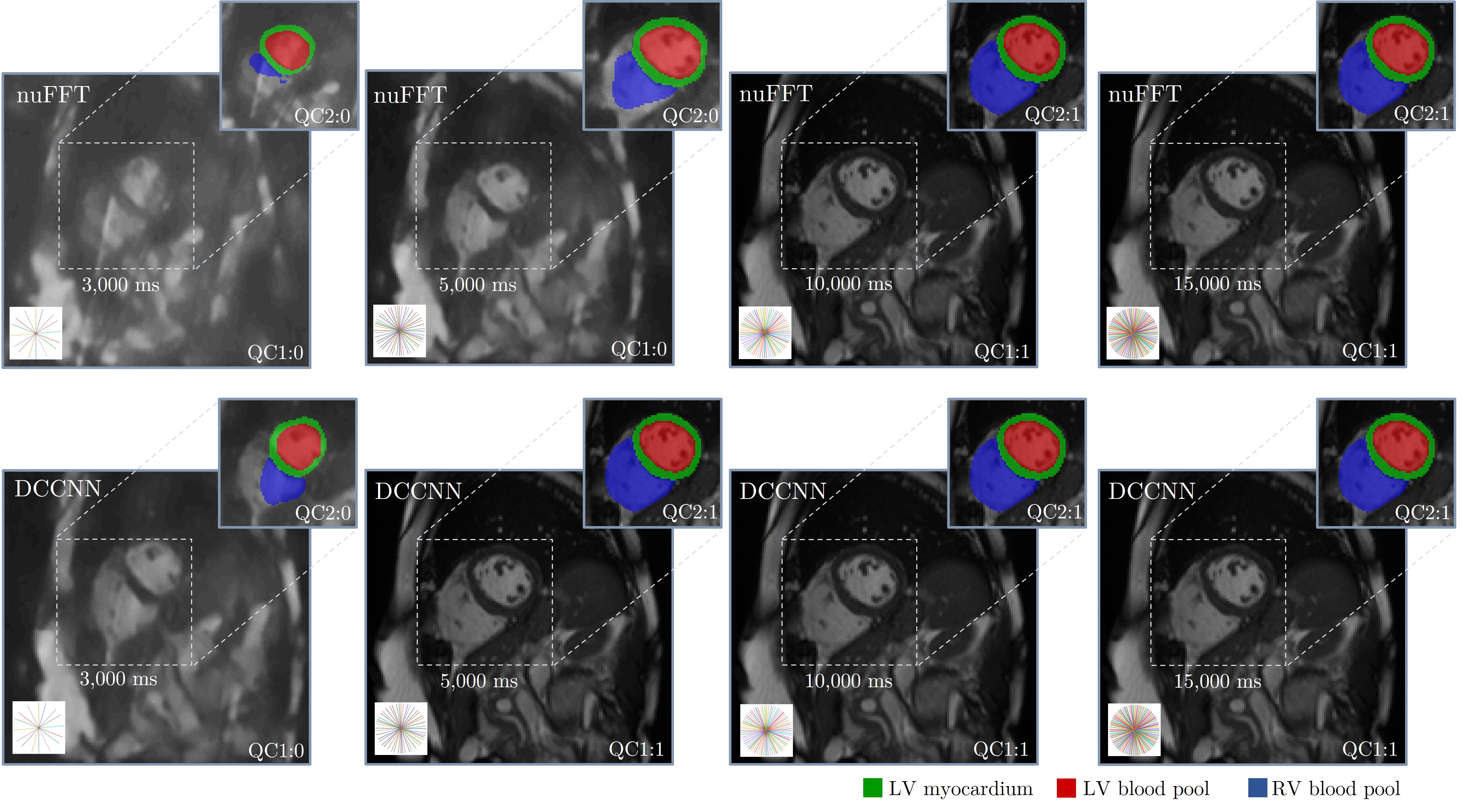}}
\caption{Illustration of image reconstructions, segmentations and undersampling trajectories as a function of the scan time using nuFFT (top row) and DCCNN (bottom row) using in-vivo radial acquisitions. For this subject, we actively reconstructed cine CMR radially-acquired undersampled k-space data, with varying degrees of undersampling (scan times from 1000 ms to 15,000 ms), using two reconstruction methods. The two QC steps were passed at a scan time of 10 seconds with the nuFFT and 4 seconds with the DCCNN, showing the ability of the framework to produce high-quality reconstructions and segmentations in a reduced scan time. QC=1 means that the QC check was passed. nuFFT: non-uniform Fast Fourier Transform. DCCNN: Deep Cascade of Convolutional Neural Networks. QC1: Image quality-control. QC2: Segmentation quality-control. ST: Scan Time.}.
\label{fig:reconstructions}
\end{figure*}

Image quality was evaluated with MAE, PSNR and SSIM, described in Section \ref{sec:metrics}, and calculated between the fully-sampled image and the undersampled image that passed QC1 with the lowest scan time during active acquisition. Table \ref{table:qc1_results} shows these metrics for Experiments I and II, described in Sections \ref{sec:simulated_data} and \ref{sec:real_data}, respectively. We show these results for the two reconstruction methods and also for healthy subjects and cardiomyopathy patients. Results are similar between healthy and cardiomyopathy patients and between nuFFT and DCCNN reconstructions and comparable with those reported in \cite{oksuz2020deep} for cine CMR image quality showing the ability of the QC check to detect high-quality reconstructions across different reconstruction algorithms and healthy/disease cases. The QC1 ResNet balanced accuracy (BACC), sensitivity (SEN) and specificity (SPE) on the testing set were equal to 92.5\%, 90\% and 95\%, respectively, for the nuFFT and 94\%, 91\% and 97\%, respectively, for the DCCNN. Fig. \ref{fig:reconstructions} illustrates cine CMR image reconstructions, segmentations and undersampling trajectories as a function of the scan time using nuFFT and DCCNN, and the output of each QC check for one healthy subject from Experiment II.
\vspace{-0.4cm}
\begin{table}[ht]
\caption{Image quality evaluated with MAE, PSNR and SSIM in 200 healthy subjects and 70 cardiomyopathy (disease) patients from the UK Biobank (Experiment I) and in-vivo acquisitions from 16 healthy subjects (Experiment II) after passing QC1. The mean and standard deviation are reported.} 
\setlength{\tabcolsep}{0.35\tabcolsep}
\centering 
\begin{tabular}{llccccc} %
\toprule
& \multirow{2}{*}{
\parbox[c]{.2\linewidth}{}}
  & \multicolumn{2}{c}{nuFFT} &&
\multicolumn{2}{c}{DCCNN} \\ 
\cmidrule{3-4} \cmidrule{6-7}
 & & {\centering Healthy} & {Disease} && {Healthy} & {Disease}  \\
\midrule
\multirow{3}*{Exp. I} & MAE & $0.03 \pm 0.03$ & $0.05 \pm 0.04$ && $0.02 \pm 0.04$ & $0.04 \pm 0.03$\\
& PSNR & $30.55 \pm 0.06$ & $29.82 \pm 0.06$ && $31.10 \pm 0.06$ & $30.09 \pm 0.07$\\
& SSIM & $0.87 \pm 0.04$ & $0.87 \pm 0.05$ && $0.91 \pm 0.03$ & $0.90 \pm 0.04$ \\
\midrule
\multirow{3}*{Exp. II} & MAE &  $0.04 \pm 0.03$ & N/A && $0.03 \pm 0.04$ & N/A\\
& PSNR & $30.82 \pm 0.05$ & N/A && $32.07 \pm 0.06$ & N/A\\
& SSIM & $0.89 \pm 0.04$ & N/A && $0.91 \pm 0.03$ & N/A\\
\bottomrule
\end{tabular}
\label{table:qc1_results}
\end{table}

\subsection{Segmentation quality analysis}
\label{sec:qc2_results}

Segmentation quality was quantified using the DSC between the GT segmentations from the fully-sampled image and segmentations that passed the QC check with the lowest scan time during active acquisition. Table \ref{table:qc2_results} shows these results for Experiments I and II, described in Sections \ref{sec:simulated_data} and \ref{sec:real_data}, respectively. We also present results for the two reconstruction methods and for healthy subjects and cardiomyopathy patients. Results are similar between healthy and cardiomyopathy patients and between nuFFT and DCCNN reconstructions showing the ability of the QC checks to detect high-quality segmentations across different reconstruction algorithms and healthy/disease cases. The QC2 ResNet BACC, SEN and SPE on the testing set were equal to 88\%, 82\% and 94\%, respectively, for the nuFFT and 88.5\%, 82\% and 95\%, respectively, for the DCCNN. 
\vspace{-0.3cm}
\begin{table}[ht]
\caption{DSC between automated and manual segmentations in 200 healthy subjects and 70 cardiomyopathy (disease) patients from the UK Biobank (Experiment I) and in-vivo acquisitions from 16 healthy subjects (Experiment II) after passing all QC checks. The mean and standard deviation are reported. LV: Left ventricle. MYO: Myocardium. RV: Right ventricle.} 
\setlength{\tabcolsep}{0.4\tabcolsep}
\centering 
\begin{tabular}{llccccc} %
\toprule
& \multirow{2}{*}{
\parbox[c]{.2\linewidth}{\centering }}
  & \multicolumn{2}{c}{nuFFT} &&
\multicolumn{2}{c}{DCCNN} \\ 
\cmidrule{3-4} \cmidrule{6-7}
& & {\centering Healthy} & {Disease} && {Healthy} & {Disease}  \\
\midrule
\multirow{3}*{Exp. I} & LV  & $0.96 \pm 0.06$  & $0.95 \pm 0.05$ &&  $0.97 \pm 0.04$  & $0.96 \pm 0.05$\\
& RV &  $0.95 \pm 0.06$ & $0.94 \pm 0.03$ &&  $0.96 \pm 0.05$ &$0.95 \pm 0.04$  \\ 
& MYO &  $0.91 \pm 0.04$ & $0.88 \pm 0.07$ &&  $0.93 \pm 0.07$ &  $0.88 \pm 0.08$ \\ 
\midrule
\multirow{3}*{Exp. II} & LV  & $0.96 \pm 0.04$  & N/A &&  $0.96 \pm 0.05$  & N/A\\
& RV &  $0.94 \pm 0.03$ & N/A &&  $0.95 \pm 0.05$ & N/A  \\ 
& MYO &  $0.91 \pm 0.04$ & N/A && $0.92 \pm 0.04$ &  N/A \\
\bottomrule
\end{tabular}
\label{table:qc2_results}
\end{table}

\begin{table*}[h!]
\caption{Experiment I: The difference in clinical measures between automated and GT segmentations for a set of 200 healthy subjects and 70 cardiomyopathy (disease) patients using two reconstruction methods: nuFFT and DCCNN. The minimum and maximum (in parenthesis), mean and standard deviation are reported.} 
\centering 
\begin{tabular}{llccccc} %
\toprule
 & \multirow{2}{*}{
\parbox[c]{.1\linewidth}{}}
  & \multicolumn{2}{c}{Healthy subjects} &&
\multicolumn{2}{c}{Cardiomyopathy patients} \\ 
\cmidrule{3-4} \cmidrule{6-7}
& & {\centering Absolute Difference} & {Relative Difference (\%)} && {Absolute Difference} & {Relative Difference (\%)}  \\
\midrule
\multirow{6}*{nuFFT} & LVEDV (ml) & $4.53 \pm 2.01$ $(0.00-5.32)$ & $3.08 \pm 2.50$ $(0.00-4.64)$ && $5.06 \pm 2.88$ $(0.07-5.32)$ & $4.03 \pm 2.60$ $(0.03-4.54)$ \\
& LVESD (ml) & $4.16 \pm 2.94$ $(0.00-6.12)$  & $4.80 \pm 2.50$ $(0.00-6.00)$ && $3.03 \pm 2.96$ $(0.01-3.40)$   & $4.48 \pm 2.75$ $(0.02-5.72)$  \\ 
& LVEF (\%) &  $2.98 \pm 2.50$ $(0.00-5.42)$ & $4.59 \pm 2.98$ $(0.01-5.68)$ && $3.07 \pm 2.75$ $(0.02-5.43)$  & $4.98 \pm 2.57$ $(0.03-5.24)$ \\ 
& RVEDV (ml) & $6.01 \pm 2.98$ $(0.23-7.24)$  & $5.99 \pm 2.50$ $(0.27-6.87)$ && $3.67 \pm 4.52$ $(0.13-4.50)$  & $3.77 \pm 3.47$ $(0.08-3.91)$ \\
& RVESV (ml) & $3.72 \pm 3.23$ $(0.16-4.45)$   & $3.18 \pm 3.51$ $(0.28-4.54)$ &&  $3.90 \pm 3.63$ $(0.14-4.75)$ & $3.96 \pm 2.96$ $(0.23-4.70)$ \\
& RVEF (\%) & $3.64 \pm 2.38$ $(0.00-5.93)$  & $7.03 \pm 4.48$ $(0.00-8.19)$ &&  $4.55 \pm 3.33$ $(0.00-4.88)$ & $ 5.91 \pm 3.86$ $(0.01-6.34)$ \\ 
\midrule
\multirow{6}*{DCCNN} & LVEDV (ml) & $2.78 \pm 2.50$ $(0.16-5.35)$ & $2.03 \pm 2.07$ $(0.13-6.14)$ && $3.62 \pm 2.48$ $(0.03-3.96)$ & $4.45 \pm 2.07$ $(0.02-4.49)$ \\
& LVESD (ml) & $3.05 \pm 2.32$ $(0.05-6.08)$  & $3.25 \pm 4.69$ $(0.09-7.00)$ && $2.19 \pm 2.51$ $(0.02-3.08)$ & $3.19 \pm 2.72$ $(0.02-4.87)$  \\ 
& LVEF (\%) &  $2.47 \pm 2.06$ $(0.01-5.67)$& $2.78 \pm 3.25$ $(0.01-6.49)$ &&  $2.91 \pm 2.61$ $(0.04-3.42)$ & $3.47 \pm 2.19$ $(0.07-4.66)$ \\ 
& RVEDV (ml) & $4.98 \pm 2.40$ $(0.01-5.03)$  &  $4.35 \pm 3.33$ $(0.01-6.42)$ && $4.01 \pm 2.20$ $(1.78-4.33)$  & $3.06 \pm 2.21$ $(1.41-3.64)$ \\
& RVESV (ml) & $2.04 \pm 3.61$ $(0.05-2.76)$   & $4.28 \pm 3.09$ $(0.07-5.08)$  && $3.07 \pm 2.92$ $(0.04-3.23)$  & $2.89 \pm 3.00$ $(0.05-3.12)$  \\
& RVEF (\%) & $2.85 \pm 4.20$ $(0.00-3.11)$ & $4.08 \pm 2.01$ $(0.01-4.82)$  && $3.06 \pm 2.17$ $(0.00-3.87)$  & $3.17 \pm 3.53$ $(0.01-4.93)$  \\ 
\bottomrule
\end{tabular}
\label{table:diff_nufft_dccnn}
\end{table*}
\vspace{-0.5cm}
\subsection{Functional parameter analysis}
\label{sec:functional_results}

\begin{figure*}[t!] 
\centering
\includegraphics[width=0.90\linewidth]{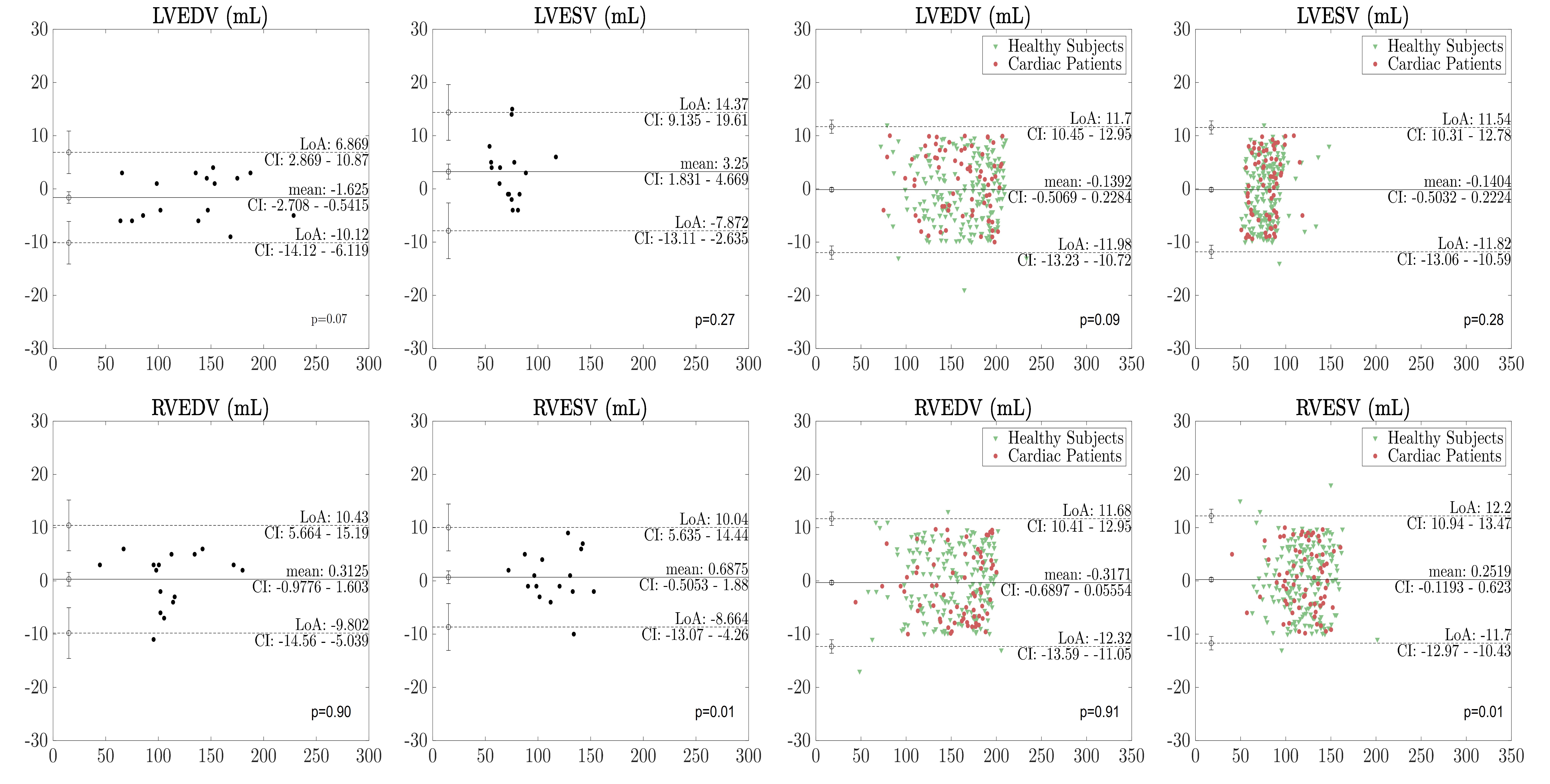}
\caption{Bland-Altman plots for the agreement between cardiac parameters estimated from fully-sampled data and via our DCCNN-based pipeline for 200 healthy subjects (in green) and 70 cardiomyopathy cases (in red), on the right-hand side, and in-vivo acquisitions of 16 healthy subjects (in black), on the left-hand side. The black solid line represents the mean bias and the black dotted lines the limits of agreement. The limits of agreement are defined as the mean difference ± 1.96 SD of differences. The error bars represent the 95\% confidence interval for both the upper and lower limits of agreement. The $p$ values represent the difference in mean bias from zero using a paired t-test.}
\label{fig:blandplots}
\end{figure*}

The performance of the proposed framework was evaluated using clinically relevant functional parameters: LVEDV, LVESV, LVEF, RVEDV, RVESV and RVEF, described in Section \ref{sec:model_estimates}. A Bland-Altman analysis for the agreement between cardiac parameters estimated from fully-sampled data and via our DCCNN-based pipeline is shown in Fig. \ref{fig:blandplots}. To verify the
significance of the biases, paired t-tests versus zero values were applied. The Pearson's correlation coefficients for Experiment I were equal to 0.98, 0.97 and 0.98 for LVEDV, LVESV and LVEF, respectively and 0.97, 0.95 and 0.96 for RVEDV, RVESV and RVEF, respectively. For Experiment II, the coefficients were equal to 0.97, 0.96 and 0.97 for LVEDV, LVESV and LVEF, respectively and 0.97, 0.95 and 0.96 for RVEDV, RVESV and RVEF, respectively. There was no significant difference in mean absolute error between cardiac patients and healthy volunteers for the output parameters.

Two measures were used to examine differences, namely the absolute and relative differences. The absolute difference is the actual difference between the predicted value and the reference value. The relative difference describes the size of the absolute difference as a fraction of the reference value. Tables \ref{table:diff_nufft_dccnn} and \ref{table:diff_16} show the mean absolute difference and mean relative difference across all subjects in clinical measures between automated and GT segmentations for Experiment I and Experiment II, respectively. Mean absolute and relative differences are within the range of intra- and inter-observer variability compared to \cite{bai2018automated}. Image quality is sufficient to allow clinically relevant parameters to be automatically estimated to within 5\% mean absolute difference. 

\begin{table}[ht]
\caption{Experiment II: The difference in clinical measures between automated segmentation and GT segmentation for 16 in-vivo radial acquisitions. The minimum and maximum (in parenthesis), mean and standard deviation are reported.} 
\setlength{\tabcolsep}{0.50\tabcolsep}
\centering 
\begin{tabular}{lcc} %
\toprule
 & {Absolute Difference} & {Relative Difference (\%)} \\
\midrule
LVEDV (ml) & $2.98 \pm 2.38$ $(0.00-4.24)$ & $3.04 \pm 3.76$ $(0.00-4.98)$  \\
LVESD (ml)& $3.98 \pm 2.55$ $(0.01-4.59)$  & $4.30 \pm 2.50$ $(0.02-5.42)$   \\ 
LVEF (\%) &  $3.01 \pm 2.50$ $(0.01-4.42)$ & $3.58 \pm 2.98$ $(0.01-4.68)$  \\ 
RVEDV (ml) & $3.72 \pm 3.97$ $(0.23-5.24)$ & $3.00 \pm 2.50$ $(0.27-4.87)$  \\
RVESV (ml) & $4.01 \pm 3.23$ $(0.15-6.45)$   & $5.18 \pm 4.51$ $(0.28-5.43)$  \\
RVEF (\%) & $3.84 \pm 2.38$ $(0.01-5.93)$  & $4.89 \pm 3.90$ $(0.01-5.10)$ \\
\bottomrule
\end{tabular}
\label{table:diff_16}
\end{table}

\subsection{Scan time}
\label{sec:scantime}

The proposed pipeline results in a reduced scan time for 2D+time cine CMR, which takes approximately 12 seconds in our clinical protocol (spatial resolution = 1.8 x 1.8 x 8.0 mm$^3$, temporal resolution = 31.56 ms and undersampling factor = 2). Table \ref{table:scantime} shows the scan times at which the QC checks are passed for Experiments I and II. By using a DCCNN for cine CMR reconstruction, we pass QC checks after approximately 4 seconds of active acquisition, i.e. an undersampling factor of 4.5 with respect to the Cartesian fully-sampled data, compared to approximately 12 seconds when using the nuFFT.

\begin{table}[ht]
\caption{Scan time, in seconds, at which image-segmentation pairs pass QC checks for a set of 200 healthy subjects and 70 cardiomyopathy (disease) patients and 16 in-vivo radial acquisitions. The mean and standard deviation are reported.} 
\setlength{\tabcolsep}{0.50\tabcolsep}
\centering 
\begin{tabular}{lccccc} %
\toprule
\multirow{2}{*}{
\parbox[c]{.2\linewidth}{\centering }}
  & \multicolumn{2}{c}{nuFFT} &&
\multicolumn{2}{c}{DCCNN} \\ 
\cmidrule{2-3} \cmidrule{5-6}
 & {\centering Healthy} & {Disease} && {Healthy} & {Disease}  \\
\midrule
Experiment I & $12.43 \pm 1.62$   & $12.85 \pm 2.03$ &&  $4.08 \pm 1.35$ & $4.02 \pm 2.23$ \\
Experiment II & $12.89 \pm 1.81$   & N/A &&  $4.01 \pm 1.12$ & N/A \\
\bottomrule
\end{tabular}
\label{table:scantime}
\end{table}

\subsection{Qualitative assessment}
\label{sec:qualitative}
\vspace{-0.1cm}
An experienced cardiologist visually assessed the predicted segmentations for 55 test subjects. According to an in-house standard operating procedure for image analysis and experience, the cardiologist visually compared automated segmentation to manual segmentation and assessed whether the two segmentations achieved a good agreement or not. The visual assessment was performed for basal, mid-ventricular and apical slices. For mid-ventricular slices, automated segmentation was found to agree well with manual segmentation for 92.7\% of the cases by visual inspection. For basal and apical slices where the ventricular contours are more complex and thus more difficult to segment, automated segmentation was found to agree well with manual segmentation for 72.7\% and 81.8\% of the cases, respectively.

\section{Discussion and conclusion}
\label{sec:discussion_conclusion}

This work demonstrates the feasibility of a DL-based framework for automated quality-controlled reconstruction and analysis of undersampled cine SAX CMR data without a previously defined level of undersampling. This framework can jointly accelerate time-consuming cine image acquisition and cumbersome manual image analysis achieving performance comparable to human experts in fully-sampled data. 

Our results show that we can produce quality-controlled images and segmentations, as shown in Tables \ref{table:qc1_results} and \ref{table:qc2_results}, respectively, in a scan time reduced from 12 to 4 seconds per slice, as shown in Table \ref{table:scantime}, enabling reliable estimates of cardiac functional parameters within 5\% mean absolute error, as shown in Tables \ref{table:diff_nufft_dccnn} and \ref{table:diff_16}, and Fig. \ref{fig:blandplots}. The image reconstruction for each image frame took approximately 30 ms. Furthermore, on a GPU, the inference time for each network in the downstream analysis was approximately 23 ms per cardiac frame. This circumvents costly image reconstructions, enabling fast post-processing immediately following accelerated and, thus, fast cine acquisitions. These times mean that real-time application of the framework on the MR scanner is feasible. 

The incorporation of robust QC steps ensures that the outputs of the framework (images, segmentations and functional metrics) are all of diagnostic quality and errors are within the range of inter-observer variability. In an automated image analysis pipeline, this method would deliver high-quality performance at high speeds and at a large scale. The framework could also provide real-time feedback during image acquisition, indicating if an acquired image is of sufficient quality for the downstream analysis tasks. 

Even though our model achieved high performance levels, one limitation is that the real CMR k-space dataset is still relatively small (16 subjects), and all datasets were acquired at a single site on a single scanner. Therefore further work needs to be done to ensure the generalization ability of the framework. Future research will need to explore more generalizable methods for analysing a wider range of CMR images, such as multi-site images acquired from different machines, different imaging protocols and integration of automated segmentation results into diagnostic reports. Nevertheless, this work represents an important proof-of-concept for the potential of integrated frameworks for reconstruction and downstream analysis. In conclusion, we believe that the proposed approach could have great clinical utility, reducing redundancies in the CMR acquisition process whilst still providing high-quality diagnostic images and robust estimates of functional parameters.

\ifCLASSOPTIONcaptionsoff
 \newpage
\fi

\bibliographystyle{IEEEtran}
\bibliography{paper}

\end{document}